\documentclass[aps,twocolumn,superscriptaddress,showpacs,pre]{revtex4}
\usepackage{graphicx}

\begin{document}
\title{Multistability at arbitrary low optical intensities in a metallo-dielectric layered structure}

\author{A. Ciattoni}
\affiliation{Consiglio Nazionale delle Ricerche, CNR-SPIN 67100 L'Aquila, Italy and Dipartimento di Fisica, Universit\`{a} dell'Aquila, 67100
             L'Aquila, Italy}

\author{C. Rizza}
\affiliation{Dipartimento di Ingegneria Elettrica e dell'Informazione, Universit\`{a} dell'Aquila, 67100 Monteluco di Roio (L'Aquila), Italy}

\author{E. Palange}
\affiliation{Dipartimento di Ingegneria Elettrica e dell'Informazione, Universit\`{a} dell'Aquila, 67100 Monteluco di Roio (L'Aquila), Italy}

\date{\today}

\begin{abstract}
We show that a nonlinear metallo-dielectric layered slab of subwavelength thickness and very small average dielectric permittivity displays optical
multistable behavior at arbitrary low optical intensities.  This is due to the fact that, in the presence of the small linear permittivity, one of
the multiple electromagnetic slab states exists no matter how small is the transmitted optical intensity. We prove that multiple states at ultra-low
optical intensities can be reached only by simultaneously operating on the incident optical intensity and incidence angle. By performing full wave
simulations, we prove that the predicted phenomenology is feasible and very robust.
\end{abstract}
\pacs{78.67.Pt, 42.65.Tg}

\maketitle

Hysteresis is probably one of the most intriguing feature a nonlinear system can exhibit both theoretically and for its main application, i.e.
designing of memory devices. In optics, bistability and related hysteresis have attracted a large research interest in the last decades
\cite{SzokeD,Chenn1} since they are the basic ingredients for devising optical memories, logic gates and optical computing devices \cite{Abraha}.
However, the actual exploitation of the hysteresis behavior has been hampered by the fact that the electromagnetic bistable phenomenology generally
occurs at high optical intensities. The research effort aimed at reducing the bistability threshold is considerable and almost all considered
strategies rely on the enhancement of nonlinear effects \cite{Stroud,Neeves,Sipeee,Fische}. Metallo-dielectric multilayer structures have been shown
to host considerable nonlinearity enhancement \cite{Bennin,Lepesh} and this has lead to the prediction of bistability \cite{Noskov,Husako}
characterized by a low intensity threshold \cite{Chennn}. Optical bistability at relatively low optical intensities has also been predicted in the
presence of surface plasmon polaritons \cite{Hicker} where the low bistability threshold is due to the strong plasmon field enhancement which in turn
produces an efficient enhancement of the nonlinear effects. Following a very different route, it has recently been proposed that an extremely marked
nonlinear behavior can be observed, instead of by enhancing the nonlinear response, by substantially reducing the medium linear dielectric
permittivity \cite{Ciatt1} and this strategy has allowed to predict low-intensity transmissivity directional hysteresis exhibited by nonlinear
metamaterial slabs with very small linear permittivity \cite{Ciatt2}.

\begin{figure}
\includegraphics[width=0.45\textwidth]{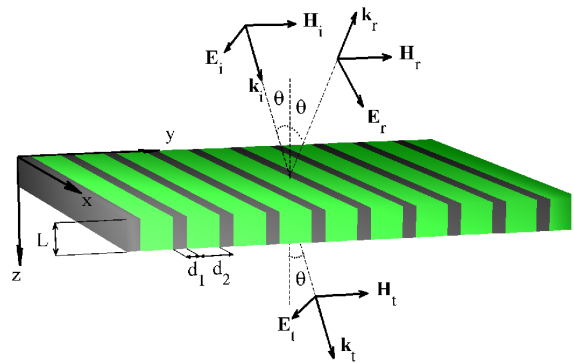}
\caption{(Color online) Geometry of the layered metallo-dielectric slab and of the TM incident (i), reflected (r) and transmitted (t) plane waves.}
\end{figure}

In this Letter, we show that a nonlinear metallo-dielectric layered slab of subwavelength thickness and very small average dielectric permittivity is
characterized by a multistable behavior which is accessible at arbitrary low optical intensities. More precisely, we prove that the slab
transmissivity, in addition to the angular multistable behavior occurring at low (but not arbitrary small) optical intensities (as discussed in
Ref.\cite{Ciatt2}), is a multi-valued function of the incident optical intensity at each fixed incidence angle. The nonlinear electromagnetic
matching at the output free-space slab interface combined with the small linear permittivity produces the multiplicity of electromagnetic states, one
of which has a large electric field component normal to the slab even if the free-space transmitted intensity approaches zero. As a consequence, the
lowest intensity at which transmissivity is multi-valued turns out to vanish for normal incidence and this implies that, at a very small incidence
angle, multistability occurs at ultra-low optical intensities. The overall slab transmissivity turns out to be a rather articulated multi-valued
function of both incident optical intensity and incidence angle and we show that ultra-low intensity multistable states can be reached only by
simultaneously varying the incident optical intensity and incidence angle along suitable paths. We perform full wave simulations to investigate both
the nontrivial metamaterial homogenization in the strong nonlinear regime and the role played by plasmonic resonances. We show that the induced
plasmon polaritons excited at the metal-dielectric interfaces affect the numerical value of the slab transmissivity solely in proximity of the
hysteresis jumps without altering the overall hysteresis behavior which is therefore proven to be a feasible and robust phenomenology.

\begin{figure}
\includegraphics[width=0.45\textwidth]{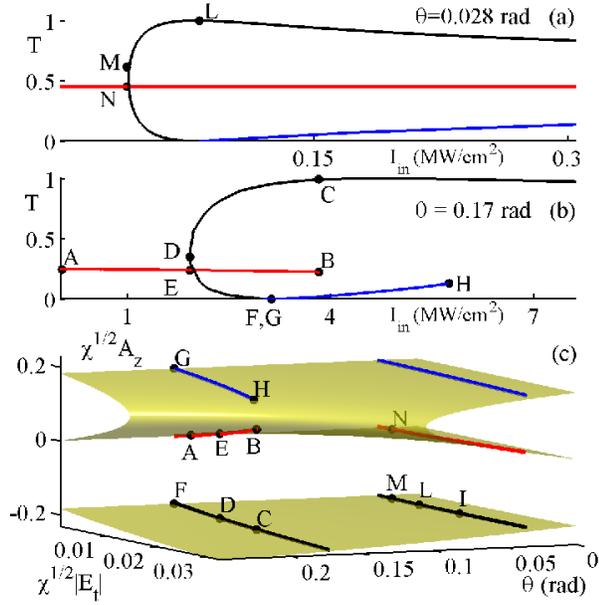}
\caption{(Color online) Nonlinear slab transmissivity $T$ (solid line) as a function of the normalized input field intensity $I_{in}$ at the fixed
incident angles $\theta = 0.028$ rad (panel (a)) and $\theta = 0.170$ rad (panel (b)). (c) Surface $|\chi|^{1/2} A_z$ and values of $|\chi|^{1/2}
A_z$ (solid lines) corresponding to the transmissivities of panels (a) and (b). In the three panels, capital letters label some reference states.}
\end{figure}

Consider the metallo-dielectric layered slab reported in Fig.1, embedded in vacuum, of thickness $L=258 \: \: nm$ and consisting of alternating,
along the $y$-axis, two isotropic and non-magnetic nonlinear Kerr layers of thicknesses $d_1$ and $d_2$. We choose to operate with a monochromatic
radiation of wavelength $\lambda = 810 \: \: nm$ and, if the layers thicknesses are such that $d_1 + d_2 \ll \lambda$, the slab behaves as an
homogeneous medium which, in the presence of a transverse magnetic field (TM), is characterized by the standard Kerr constitutive relations ${\bf D}
= \epsilon_0 \left\{ \epsilon {\bf E} + \chi \left[\left({\bf E \cdot E^*} \right) {\bf E} + \frac{1}{2} \left({\bf E \cdot E} \right) {\bf E}^*
\right] \right\}$ \cite{Ciatt1}. The effective dielectric permittivity $\epsilon$ and nonlinear susceptibility $\chi$ are the average values of their
underlying constituents and we choose here $\epsilon = -0.054$ and $\chi = 8.72 \times 10^{-17} \: m^2/V^2$ (see below for the actual feasibility of
these values). Note that the negative and very small value of the effective dielectric permittivity is achieved here by averaging the negative and
positive permittivities characterizing metal and dielectric layers, respectively. As reported in Fig.1, a TM plane wave of amplitude $E_i$ is made to
impinge onto the slab interface at $z=0$ with incidence angle $\theta$ thus producing a TM reflected plane wave of amplitude $E_r$ and a transmitted
TM plane wave of amplitude $E_t$. We have numerically evaluated the slab transmissivity $T=|E_t/E_i|^2$ for a number of incident optical intensities
and incidence angles and in Fig.2 we report $T$ as a function of the input intensity $I_{in} = (1/2) \sqrt{\epsilon_0 / \mu_0} |E_i|^2$ for the
incidence angles $\theta = 0.028$ rad (panel (a)) and $\theta = 0.170$ rad (panel (b)). The reported slab transmissivities are multi-valued functions
of the incident optical intensity. The multiplicity (at last three-fold) of different slab electromagnetic states, corresponding to the same
transmissivity, physically arises (as discussed in Ref.\cite{Ciatt2}) from the impact of the nonlinearity on the electromagnetic matching condition
at the slab interfaces, an effect made possible by the very small value of $|\epsilon|$. Specifically, the continuity, at $z=L$, of the displacement
field component normal to the output slab interface, $D_z(L^-)=D_z(L^+)$, yields $\left[\epsilon + \frac{3}{2} \chi \left( |E_t|^2 \cos^2 \theta +
A_z^2 \right) \right] A_z+ |E_t| \sin \theta = 0$, which is a cubic equation that, for a given $E_t$ and $\theta$ yields at last three different
values of the real amplitude $A_z =( E_t/|E_t| ) \exp( -i \cos \theta) E_z(L)$ characterizing the longitudinal electric field component at $z=L$. In
Fig. 2(c) we report the surface comprising all the possible values of $|\chi|^{1/2} A_z$ together with values of $|\chi|^{1/2} A_z$ (solid lines)
corresponding to the transmissivities of Fig.2(a) and (b), the capital letters labelling some reference states. Note that the incident intensities at
which multistability occurs (of the order of $0.15 \: \: MW/cm^2$ and $4 \: \: MW/cm^2$ for the transmissivities of Fig.2(a) and 2(b), respectively)
are smaller than those normally required for observing standard optical bistability. However it is evident that the smaller $|\epsilon|$ the smaller
the intensity threshold required for observing the considered multistability so that very smaller intensity thresholds (of the order of $W/cm^2$) are
very likely to be attained.

Multistable transmissivities of Fig.2(a) and 2(b) show two main mutual differences which deserve to be discussed. First, multi-valued transmissivity
of Fig.2(a) has, for $I_{in} > I_N$, three different branches that does not break whereas multi-valued transmissivity of Fig.2(b) has, for $I_{in} >
I_E$, an upper branch that does not break and two lower branches characterized by the two breaking states $B$ and $H$. The origin of this difference
is easily understood by considering the surface of Fig.2(c) from which it is evident that the breaking of a transmissivity branch occurs for states
close to critical points at which surface folding takes place (states $B$ and $H$ of Fig.2(c)). Second, even though in both cases there exists a
lower intensity above which the transmissivity is multi-valued (see the state N and E of Fig.2(a) and 2(b), respectively), transmissivities of
Fig.2(a) and 2(b) are multistable for $I_{in} > I_N = 0.04 \: \: MW/cm^2$ and $I_{in} > I_E = 1.92 \: \: MW/cm^2$, respectively so that $I_{N} \ll
I_{E}$ or, in other words the smaller the incidence angle the smaller the incident optical intensity at which multistable states are accessible. In
order to closer investigate this remarkable point we have plotted in Fig.3 the slab transmissivity as a function of both the incident optical
intensity and incidence angle thus providing a complete description of the slab multistability. The transmissivity of Fig.3 is a multi-sheet surface
along which we have also plotted, for clarity purposes, the transmissivities of Fig.2(a) and 2(b) (dashed lines) together with the very same
reference states labelled with capital letters. The intersection between the upper and intermediate sheets yields the lower intensity at which
multistability occurs and it turns out that such intensity is a monothonically increasing function of incidence angle $\theta$ and, most importantly,
that this intensity vanishes if the angle vanishes. To underline this result, in the inset of Fig.3 we have reported the region of the independent
external parameter plane $(I_{in},\theta)$ where multistability occurs (shaded region) and, since this region contains the excitation state
$I_{in}=0$ $\theta=0$, it is evident that multistable electromagnetic slab states are available at arbitrary small incident optical intensities (and
very small incidence angles). In order to physically grasp the origin of such an unusual situation, consider the states providing multistability and
belonging to the lower sheet of the surface reported in Fig.2(c). For $|E_t|=0$ we have $A_z =0$ and $A_z = \pm \sqrt{-2\epsilon / (3\chi)} = \pm
0.19 / \sqrt{\chi}$ (see the above mentioned matching condition at $z=L$) so that the surface exists even at arbitrary small values of $|E_t|$ (see
Fig.2(c)) and multistable states exist even in the limiting case where there is no transmitted waves. Since the input intensity is smaller than or
equal to the output one, it is evident that the described mechanism provides multistable states even at ultra-low input optical intensities.

\begin{figure}
\includegraphics[width=0.45\textwidth]{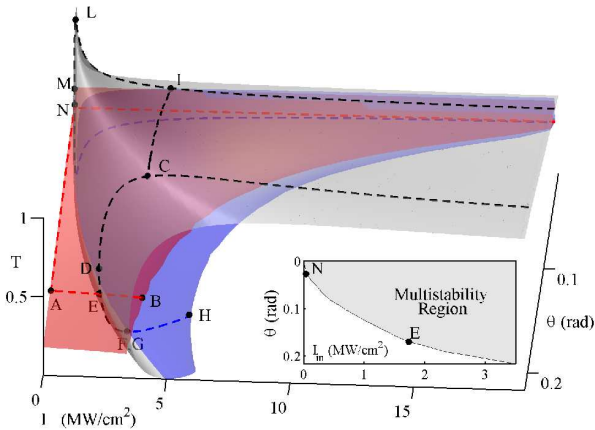}
\caption{(Color online) Slab transmissivity $T$ as a function of both the incident optical intensity $I_{in}$ and incidence angle $\theta$. Dashed
lines represent the slab transmissivity of Fig.2(a) and 2(b) and the capital letters label the same reference states as in Fig.2. In the inset the
plane of independent parameters $I_{in}$ and $\theta$ is reported together with the region at which slab multistability occurs (shaded region).}
\end{figure}

The transmissivity reported in Fig.2(b) clearly shows hysteresis behavior since the state $B$ is a breaking point of its transmissivity branch.
Suppose to switch on the incident plane wave with $\theta = 0.17$ rad and $I_{in} = I_A = 0.035 MW/cm^2$ so that the state A of Fig.2(b), 2(c) and
Fig.3 is excited since before the illumination the slab was not polarized. By increasing the input intensity and holding the incidence angle fixed,
the transmissivity follows the curve joining the point A and B of Fig.2(b) while the electromagnetic state continuously varies as in Fig. 2(c). From
the state B, if the intensity is further increased, it is evident that the electromagnetic state undergoes a sudden jump to the state C (see
Fig.2(c)) on the lower surface sheet since there is no allowed state continuously joined to B. As a consequence, the transmissivity undergoes a
sudden jump to a higher value (see Fig.2(b)). If now, starting from the state C, the input intensity is decreased, the electromagnetic state varies
along the curve from C to D of Fig.2(c) whereas the transmissivity in Fig.2(b) assumes the values along the curve from C to D which are different
from those attained along the forward path (hysteresis). At the state D, if the intensity is further decreased, the state undergoes a jump to the
state E of Fig.2(c), belonging to the central surface sheet, since the states from D to F of Fig.2(c) have, according to Fig.2(b), higher input
intensities.

The transmissivity reported if Fig.2(a), although multi-valued, does not exhibit hysteresis since its branches do not present breaking points.
However the low intensity multistable states can always be reached by selecting appropriate paths of the excitation plane $(I_{in},\theta)$. Suppose,
for example, the multistability responsible state $L$ of Fig.2(a), 2(c) and Fig.3 (of intensity $I_L = 0.08 \: \: MW/cm^2$) is required to be
reached. A possible operational way consists in exciting the state A and reaching the state C as discussed in the above hysteresis example. Now if
from the state $C$ the intensity is hold fixed and the angle is decreased, from Fig.3 it is evident that the state I is reached since both the states
C and I belongs to the same surface sheet of Fig.2(c). From the state C one can now fix the angle and decrease the incident optical intensity thus
eventually reaching the prescribed target state L. From L one can decrease the intensity thus reaching the state $M$ from which, a further intensity
decrement, produces (in analogy to the above described hysteresis loop) a sudden jump of the trasmittivity to $N$ of Fig.2(a), the corresponding
state $N$ of Fig.2(c) belonging to the central surface sheet. The very same procedure can even be used to excite the above discussed multistable
states at much smaller incident optical intensities (and incidence angle).

\begin{figure}
\includegraphics[width=0.45\textwidth]{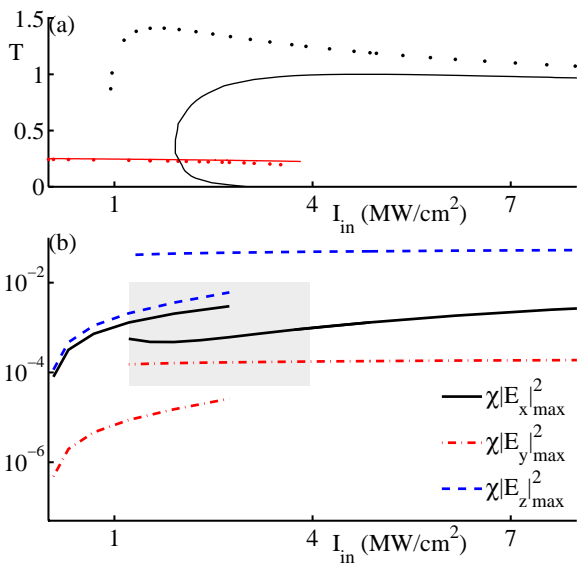}
\caption{(Color online) (a) Comparison between the slab transmissivities evaluated through full-wave simulations (dotted lines) and those of Fig.2(b)
 (solid lines). (b) Semi-log plot of the maximum values, within the layered medium bulk, of the normalized squared field components as a function
of the normalized input field amplitude obtained for the full-wave evaluation of the slab transmissivity reported in panel (a).}
\end{figure}

In order to discuss the feasibility of the above predicted slab multistability we have performed 3D full-wave finite-element simulations
\cite{Comsol} for evaluating the transmissivity of the slab in the presence of a TM radiation of free-space wavelength $\lambda = 810$ nm. We have
chosen $d_1 = 2$ nm, $d_2 = 5.25$ nm, $L = 258$ nm, $\epsilon_1 = -28.81 + 10 i$, $\epsilon_2 = 10.9 -3.8 i$, $\chi_1 = 3.16 \times 10^{-16} \:
m^2/V^2$ and $\chi_2 = 0$. The parameters of medium $1$ coincide with those of silver \cite{Palikk}, characterized by a very large nonlinear
susceptibility \cite{YangGu}, with the imaginary part of the permittivity corrected by the layer size effect (since $d_1 = 2$ nm) \cite{CaiSha}
whereas medium $2$ is a linear dielectric with gain (to compensate the metal losses). The effective permittivity and nonlinear susceptibility of the
considered sample are $\epsilon = (d_1 \epsilon_1 + d_2 \epsilon_2) / (d_1 + d_2) = -0.054 + 0.007i$ and $\chi = (d_1 \chi_1 + d_2 \chi_2) / (d_1 +
d_2) = 8.72 \times 10^{-17} \: m^2/V^2$. In Fig.4(a) we report the comparison between the transmissivity evaluated through full-wave simulations for
$\theta = 0.17$ rad (dotted line) and the transmissivity of Fig.2(b). We note that good agreement exists between the results of the two kind of
simulations and, most importantly, that finite-element simulations still predicts the above discussed multistability thus proving its robustness. The
origin of the discrepancies between the transmissivities lies in the fact that a surface plasmon resonance occurs when the TM wave impinges on the
slab as reported in Fig.1. Therefore an electric field $y$-component ($E_y$) of plasmonic origin arises mainly at the edges of the layers inside the
medium and it is characterized by a sub-wavelength varying profile and an evanescent field tail in vacuum which does not contribute to the power
flow. Evidently, if $E_y$ is much smaller than $E_x$ and $E_z$, the homogenization theory correctly describes the slab nonlinear behavior. In
Fig.4(b) we report the maximum values (within the medium) attained by the three field components for $\theta = 0.17$ rad as a function of the
incident optical intensity corresponding to the evaluated full-wave transmissivities reported in Fig.4(a). It is evident that, outside the shaded
region, $E_y$ (dot-dashed line) is much smaller than both $E_x$ and $E_z$ (solid and dashed lines respectively) and in the corresponding regions of
Fig.4(a) the agreement with the homogenization approach is very satisfactory. Within the shaded region of Fig.4(b), $E_y$ is not negligible and this
partially breaks the validity of the homogenization approach thus leading to the discrepancies of the transmissivities in the range $1 \: \: MW/cm^2
< I_{in} < 4 \: \: MW/cm^2$ as reported in Fig.4(a).

In conclusion we have shown that a nonlinear metallo-dielectric layered slab of subwavelength thickness with very small average permittivity is
characterized by a very low bistability threshold and, most importantly, that it provides multistable states at arbitrary low optical intensities.
Such ultra-low intensity multistable states are always accessible through a suitable excitation procedure involving both the incident intensity and
the incidence angle. Since the main difficulty of usefully exploiting standard optical bistability is related to the required high values of the
optical intensity, we believe that our findings can have a large impact on the devising of a novel class of nanophotonic devices with complex memory
functionalities and operating at very low optical intensities (as compared to those generally required in standard nonlinear optics).



\begin{thebibliography} {aa}

\bibitem{SzokeD} A. Szoke, V. Daneu, J. Goldhar and N. A. Kurnit, Appl. Phys. Lett. \textbf{15}, 376 (1969).
\bibitem{Chenn1} W. Chen and D. L. Mills, Phys. Rev. B \textbf{35}, 524 (1987).
\bibitem{Abraha} E. Abrhaham and S. D. Smith, Rep. Prog. Phys. \textbf{45}, 815 (1982).
\bibitem{Stroud} D. Stroud and Van E. Wood, J. Opt. Soc. Am. B \textbf{6}, 778 (1989).
\bibitem{Neeves} A. E. Neeves and M. H. Birnboim, J. Opt. Soc. Am. B \textbf{6}, 787 (1989).
\bibitem{Sipeee} J. E. Sipe and R. W. Boyd, Phys. Rev. A \textbf{46}, 1614 (1992).
\bibitem{Fische} G. L. Fischer, R. W. Boyd, R. J. Gehr, S. A. Jenekhe, J. A. Osaheni, J. E. Sipe and L. A. Weller-Brophy,
                 Phys. Rev. Lett. \textbf{74}, 1871 (1995).
\bibitem{Bennin} R. S. Bennink, Y. Yoon, R. W. Boyd and J. E. Sipe, Opt. Lett \textbf{24}, 1416 (1999).
\bibitem{Lepesh} N. N. Lepeshkin, A. Schweinsberg, G. Piredda, R. S. Bennink and R. W. Boyd, Phys. Rev. Lett \textbf{93}, 123902 (2004).
\bibitem{Noskov} R. E. Noskov and A. A. Zharov, Opto-Electron. Rev. \textbf{14}, 217 (2006).
\bibitem{Husako} A. Husakou and J. Herrmann, Phys. Rev. Lett. \textbf{99}, 127402 (2007).
\bibitem{Chennn} J. Chen, P. Wang, X. Wang, Y. Lu and R. Zheng, App. Phys. Lett. \textbf{94}, 081117 (2009).
\bibitem{Hicker} R. K. Hickernell and D. Sarid, J. Opt. Soc. Am. B \textbf{3}, 1059 (1986).
\bibitem{Ciatt1} A. Ciattoni, C. Rizza and E. Palange, Phys. Rev. A \textbf{81}, 043839 (2010).
\bibitem{Ciatt2} A. Ciattoni, C. Rizza and E. Palange, Opt. Lett. \textbf{35}, 2130 (2010).
\bibitem{Comsol} COMSOL, www.comsol.com
\bibitem{Palikk} E. D. Palik, {\it Handbook of Optical Constants of Solids} (Academic Press, San Diego, 1998).
\bibitem{YangGu} G. Yang, D. Guan, W. Wang, W. Wu, Z. Chen, Opt. Mater. \textbf{25}, 439 (2004).
\bibitem{CaiSha} W. Cai and V. Shalaev, {\it Optical Metamaterials: Fundamentals and Applications} (Springer, Dordrecht, 2010).

\end{thebibliography}
\end{document}